\documentstyle[epsf,epsfig,amssymb,secnumtab]{fbssuppl}

\newcommand{\reduction}[2]{#1 \biggr|_{#2}}
\title{Structure of T\,--\, and S\,--\,Matrices
in Unphysical Sheets and Resonances in Three\,--\,Body
Systems%
\thanks{Contribution to Proceeding of the 16th European Conference 
on Few-Body Problems in Physics, Autrans (France), 1--6~June 1998. 
LANL E-print {\tt physics/9810006}.}}
\author{A. K. Motovilov\instnr{1,2},
E. A. Kolganova\instnr{2}}
\instlist{Physikalishes Institut der Universit\"at Bonn,
      Endenicher Allee 11\,--\,13, \mbox{D-53115} Bonn, Germany
\and Joint Institute for Nuclear Research, 141980 Dubna,
Moscow region, Russia}
\begin{document}
\setcounter{volume}{10}
\setcounter{page}{75}
\maketitle

\begin{abstract}
Algorithm, based on explicit representations for the analytic
continuation of Faddeev components of the three-body T-matrix in
unphysical energy sheets, is employed to study mechanism of
disappearance and formation of the Efimov levels of the helium
$^4$He trimer.
\end{abstract}
\section{Introduction}
\label{Intro}

Explicit representations for the Faddeev components of the
three-body T-matrix continued analytically into unphysical
sheets of the energy Riemann surface have been formulated and
proved recently in Ref.~\cite{MotMathNachr}.  According to the
representations, the T-matrix in unphysical sheets is explicitly
expressed in terms of its components only taken in the physical 
sheet.  Analogous explicit representations were also found for 
the analytic continuation of the three-body scattering matrices.  
These representations disclose the structure of kernels of the 
T- and S-matrices after continuation and give new capacities for 
analytical and numerical studies of the three-body resonances.  
In particular the representations imply that the resonance poles 
of the S-matrix as well as T-matrix in an unphysical sheet 
correspond merely to the zeros of the suitably truncated 
three-body scattering matrix taken in the physical sheet. 
Therefore, one can search for resonances in a certain unphysical 
sheet staying always, nevertheless, in the physical sheet and 
only calculating the position of zeros of the appropriate 
truncation of the total three-body scattering matrix. This 
statement holds true not only for the case of the conventional 
smooth quickly decreasing interactions but also for the case of 
the singular interactions described by different variants of the 
Boundary Condition Model, in particular for the inter--particle 
interactions of a hard-core nature like in most molecular 
systems.

As a concrete application of the method, we present here the 
results of our numerical study of the simplest truncation of the 
scattering matrix in the $^4$He three-atomic system, namely of 
the $(2+1\to2+1)$ S-matrix component corresponding to the 
scattering of a $^4$He atom off a $^4$He dimer. The point is 
that there is already a series of works 
\cite{Gloeckle}--\cite{KMS-JPB} (also see 
Refs.~\cite{Nielsen}--\cite{TFDA}) showing that the excited 
state of the $^4$He trimer is initiated by the Efimov effect 
~\cite{VEfimov}.  In these works, various versions of the 
$^4$He--$^4$He potential were employed.  However, the basic 
result of Refs.~\cite{Gloeckle}--\cite{KMS-JPB} on the excited 
state of the helium trimer is the same: this state disappears 
after the interatomic potential is multiplied by the increasing 
factor $\lambda$ when it approaches the value about 1.2.  It is 
just such a nonstandard behavior of the excited-state energy as 
the coupling between helium atoms becomes more and more 
strengthening, points to the Efimov nature of the trimer excited 
state. The present work is aimed at elucidating the fate of the 
trimer excited state upon its disappearance in the physical 
sheet when $\lambda>1$ and at studying the mechanism of arising 
of new excited states when $\lambda<1$. As the interatomic 
He\,--\,He potential, we use the potential HFD-B~\cite{Aziz87}.  
We have established that for such He\,--\,He\,-\,interactions 
the trimer excited-state energy merges with the two-body 
threshold $\epsilon_d$ at $\lambda\approx1.18$ and with further 
decreasing $\lambda$ it transforms into a virtual level of the 
first order (a simple real pole of the analytic continuation of 
the $(2+1\to2+1)$ scattering matrix component) lying in the 
unphysical energy sheet adjoining the physical sheet along the 
spectral interval between $\epsilon_d$ and the three--body 
threshold.  We trace the position of this level for $\lambda$ 
increasing up to 1.5.  Besides, we have found that the excited 
(Efimov) levels for $\lambda<1$ also originate from virtual 
levels of the first order that are formed in pairs.  Before a 
pair of virtual levels appears, there occurs a fusion of a pair 
of conjugate resonances of the first order (simple complex poles 
of the analytic continuation of the scattering matrix in the 
unphysical sheet) resulting in the virtual level of the second 
order.

\section{Representations for three-body T\,--\, and
         S\,--\,matrices \newline in unphysical energy sheets}
\label{Represent}

The method used for calculation of resonances in the present
work, is based on the explicit
representations~\cite{MotMathNachr} for analytic continuation of
the T- and scattering matrices in unphysical sheets which hold
true at least for a part of the three-body Riemann surface.  To
describe this part we introduce the auxiliary vector-function
${\frak f}(z)=({\frak f}_{0}(z),{\frak f}_{1,1}(z),..., {\frak
f}_{1,n_1}(z),$ ${\frak f}_{2,1}(z),...,{\frak f}_{2,n_2}(z),$
${\frak f}_{3,1}(z),...,{\frak f}_{3,n_3}(z))$ $\enspace$ with
$\enspace{\frak f}_{0}(z)$ $ =$ $\ln{z}$ and ${\frak
f}_{\alpha,j}(z)$ $=$ $(z-\lambda_{\alpha,j})^{1/2}.$ Here, by
$z$ we understand the total three-body energy in the c.\,m.
system and by $\lambda_{\alpha,j}$, the respective binding
energies of the two-body subsystems $\alpha,$ $\alpha=1,2,3,$
$j=1,2,...,n_\alpha$, $n_\alpha<\infty$.  The sheets $\rmPi_{l}$
of the Riemann surface of the vector-function ${\frak f}(z)$ are
numerated by the multi-index
$
l=( l_{0},l_{1,1},..., l_{1,n_1},
l_{2,1},...,l_{2,n_2},
l_{3,1},...,l_{3,n_3}),
$
where $l_{\alpha,j}=0$ if the sheet $\rmPi_{l}$ corresponds to the
main (arithmetic) branch of the square root
$(z-\lambda_\alpha)^{1/2}.$ Otherwise, $l_{\alpha,j}=1$ is
assumed.  Value of $l_0$ coincides with the number of the branch
of the function $\ln{z}$, $\ln z=\ln|z|+\,{\rm i}\,2\pi l_0+{\rm 
i}\phi$ where $\phi=\mathop{\rm arg}z$. For the physical sheet 
identified by $l_0=l_{\alpha,j}=0$, $\alpha=1,2,3,\enspace 
j=1,2,...,n_\alpha$, we use the notation $\rmPi_0$.

Surely, the structure of the total three-body Riemann surface is
essentially more complicated than that of the auxiliary function
${\frak f}$.  For instance, the sheets $\rmPi_{l}$ with $l_0=\pm
1$ have additional branching points corresponding to resonances
of the two-body subsystems.  The part of the total three-body
Riemann surface where the representations of
Ref.~\cite{MotMathNachr} are valid, consists of the sheets
$\rmPi_l$ of the Riemann surface of the function ${\frak f}$
identified by $l_0=0$ (such unphysical sheets are called {\it
two-body} sheets) and two {\it three-body} sheets identified by
$l_0=\pm 1$ and $l_{\alpha,j}=1,$ $\alpha=1,2,3,$
$j=1,2,...,n_\alpha.$

In what follows by $k_\alpha,p_\alpha$
($k_\alpha,p_\alpha\in{\Bbb R}^3$ or $k_\alpha,p_\alpha\in{\Bbb
C}^3$) we understand the standard reduced relative momenta of
the three-body system while \mbox{$P=\{k_\alpha,p_\alpha\}$}
($P\in{\Bbb R}^6$ or $P\in{\Bbb C}^6$) stands for the total
relative momentum.

The representations~\cite{MotMathNachr} for the analytic continuation
of the matrix
$
{\bf M}(z)=\left\{ M_{\alpha\beta}(z) \right\},
\alpha,\beta=1,2,3,
$
of the Faddeev components $M_{\alpha\beta}(z)$ (see~\cite{MF})
of the three-body $T$-operator, into the sheet $\rmPi_l$ read as
follows%
\footnote{One assumes that all the pair interactions fall off in
the coordinate space not slower than exponentially and, thus, their
Fourier transforms $v_\alpha(k),$ $k\in{\Bbb C}^3,$ are
holomorphic functions of the relative momenta $k$ in a stripe
$|\mathop{\rm Im} k|<b$ for some $b>0$.}:
\begin{equation}
\label{Mab}
\reduction{{\bf M}(z)}{\rmPi_l}= {\bf M}(z) - {\bf B}^{\dagger}(z)
A(z)L {\rm S}_{l}^{-1}(z)\widetilde{L} {\bf B}(z).
\end{equation}
Here, the factor $A(z)$ is the diagonal matrix,
$A(z)=\mathop{\rm diag}\{A_{0}(z),$ $A_{1,1}(z),$ $\ldots,$
$A_{1,n_3}(z)\}$ with $A_{0}(z)=-\pi {\rm i} z^2$ and
$A_{\alpha,j}=-\pi {\rm i} \sqrt{z-\lambda_{\alpha,j}},$
$j=1,2,...,n_\alpha$.  Notations  $L$ and $\widetilde{L}$ stand for
the diagonal number matrices combined of the indices of the
sheet $\rmPi_l$:  $L=\mathop{\rm
diag}\{l_0,l_{1,1},...,l_{3,n_3}\}$ and $\widetilde{L}=\mathop{\rm
diag}\{|l_0|, l_{1,1},...,l_{3,n_3}\}.$ By ${\rm S}_l(z)$ we
understand a truncation of the three-body scattering matrix
${\rm S}(z)$ defined in \mbox{$\widehat{\cal
G}=L_2(S^5)\mathop{\oplus}\limits_{\alpha=1}^{3}
\mathop{\oplus}\limits_{j=1}^{n_\alpha}L_2(S^2)$} by the
equation
$$
{\rm S}_l(z)=\widehat{I}+\widetilde{L}\bigl[{\rm S}(z)-\widehat{I}\bigr]L
$$
where $\widehat{I}$ is the identity operator in  $\widehat{\cal
G}$.  Also, we use the notations
$$
{\bf B}(z)=\left( \begin{array}{l}
     {\rm J}_{0}\rmOmega{\bf M}    \\
     {\rm J}_1\bm{\rmPsi}^{*}[\rmUpsilon{\bf M}+{\bf v}]
    \end{array}
\right)
\mbox{\,\,and\,\,}
{\bf B}^{\dagger}(z)=\left({\bf M}(z)\rmOmega^\dagger{\rm
J}_{0}^{\dagger}, [{\bf v} +{\bf M}\rmUpsilon]\bm{\rmPsi}{\rm
J}^\dagger_1\right).
$$
Here,
$
    {\bf v}=\mathop{\rm diag}\{v_1,v_2,v_3\}
$
with $v_\alpha$, the pair potentials, $\alpha=1,2,3$. At the
same time, $\rmOmega=(1,\,\,1,\,\,1),$
$\rmUpsilon=\{\rmUpsilon_{\alpha\beta}\}$ with
$\rmUpsilon_{\alpha\beta}=1-\delta_{\alpha\beta}$,
$\alpha,\beta=1,2,3,$ and $\bm{\rmPsi}=\mathop{\rm
diag}\{\bm{\rmPsi}_1,\bm{\rmPsi}_2,\bm{\rmPsi}_3\}$ where $\bm{\rmPsi}_\alpha,$
$\alpha=1,2,3,$ are operators acting on
$f=(f_1,f_2,...,f_{n_\alpha})\in
\mathop{\oplus}\limits_{j=1}^{n_\alpha} L_2({\Bbb R}^3)$ as
$(\bm{\rmPsi}_\alpha f)(P)=
\sum\limits_{j=1}^{n_\alpha}\psi_{\alpha,j}(k_\alpha)f_{j}(p_\alpha)$
where, in turn,  $\psi_{\alpha,j}$ is the bound-state wave
function of the pair subsystem $\alpha$ corresponding to the
binding energy $\lambda_{\alpha,j}$.  By $\bm{\rmPsi}^{*}$ we denote
operator adjoint to $\bm{\rmPsi}$.  Notation ${\rm J}_{0}(z)$ is used
for the operator restricting a function on the energy-shell
$|P|^2=z$.  The diagonal matrix-valued function ${\rm
J}_1(z)=\mathop{\rm diag}\{{\rm J}_{1,1}(z),..., {\rm
J}_{3,n_3}(z)\}$ consists of the operators ${\rm
J}_{\alpha,j}(z)$ of restriction on the energy surfaces
$|p_\alpha|^{2}=z-\lambda_{\alpha,j}.$ The operators
$\rmOmega^\dagger$, ${\rm J}_{0}^{\dagger}(z)$ and ${\rm
J}^\dagger_1(z)$ represent the ``transposed'' matrices $\rmOmega$,
${\rm J}_{0}(z)$ and ${\rm J}_1(z)$, respectively.  Operators
${\rm J}_{0}^{\dagger}(z)$ and ${\rm J}^\dagger_1(z)$  act in
the expression for ${\bf B}^{\dagger}$ (as if) to the left.

With some stipulations (see~\cite{MotMathNachr}) the
representations for the scattering matrix read
\begin{equation}
\label{SS}
\reduction{{\rm S}(z)}{\rmPi_l}={\cal E}(l)\left\{ \widehat{I} +
{\rm S}_{l}^{-1}(z) [{\rm S}(z)-\widehat{I}]e(l)\right\} {\cal E}(l).
\end{equation}
Here, $ {\cal E}=\mathop{\rm diag}\{{\cal E}_0 ,{\cal
E}_{1,1},..., {\cal E}_{3,n_3} \} $ where $ {\cal E}_0 $ is the
identity operator in $L_2 (S^5 )$ if  $l_0 =0$ and inversion,
$({\cal E}_0 f)(\widehat{P}) = f(-\widehat{P}),$ if $l_0 =\pm
1$.  Analogously, $ {\cal E}_{\alpha,j} $ is the identity
operator in $L_2 (S^2 )$ for $ l_{\alpha,j} =0 $ and inversion
for $ l_{\alpha,j} =1.$ Notation $e(l)$ is used for the diagonal
number matrix $e(l)=\mathop{\rm diag}\{e_0
,e_{1,1},...,e_{3,n_3} \}$ with nontrivial elements
$e_{\alpha,j}=1$ if $l_{\alpha,j}=0$ and $e_{\alpha,j}=-1$ if
$l_{\alpha,j}=1;$ for all the cases $e_0=1$.

It follows from the representations~(\ref{Mab}) and ({\ref{SS})
that the resonances (the nontrivial poles of $\reduction{{\bf
M}(z)}{\rmPi_l}$ and $\reduction{{\rm S}(z)}{\rmPi_l}$) situated in the
unphysical sheet $\rmPi_l$ are those points $z=z_{\rm res}$ in the
physical sheet where the matrix ${\rm S}_l(z)$ has zero as eigenvalue.
Therefore, {\em calculation of resonances in the unphysical
sheet $\rmPi_l$ is reduced to a search for zeros of the respective
truncation ${\rm S}_l(z)$ of the scattering matrix ${\rm S}(z)$ in the
physical sheet}.

\section{Method for search of resonances in a three--body system
\newline on the basis of the Faddeev differential equations}
\label{DiffFS}

In this work we discuss the example of the three-atomic $^4$He
system at the total angular momentum $L=0$.  We
consider the case where the interatomic interactions include a
hard core component and, outside the hard core domain, are
described by conventional smooth potentials.  In this case, the
angular partial analysis reduces the initial Faddeev equation
for three identical bosons to a system of coupled
two-dimensional integro-differential equations (see
Ref.~\cite{KMS-JPB} and references therein)
\begin{eqnarray}
\label{FadPartCor}
\lefteqn{   \left[-\frac{\partial^2}{\partial x^2}
            -\frac{\partial^2}{\partial y^2}
            +l(l+1)\left(\frac{1}{x^2}
            +\frac{1}{y^2}\right)
    -E\right]F_l(x,y)   }
\qquad\qquad\qquad\qquad\qquad\qquad \\
\nonumber
&&=\left\{\begin{array}{cl} -V(x)\Psi_l(x,y), & x>c \\
                    0,                  & x<c\,.
\end{array}\right.
\end{eqnarray}
Here, $x,y$ stand for the standard Jacobi variables and $c$, for
the core range.  At $L=0$ the partial angular momentum $l$
corresponds both to the dimer subsystem and a complementary
atom.  For the $S$-state three-boson system $l$ is even,
$l=0,2,4,\ldots\,.$ In our work, the energy $z$ can get both
real and complex values.  The He--He potential $V(x)$ acting
outside the core domain is assumed to be central.  The partial
wave function $\Psi_l(x,y)$ is related to the Faddeev components
$F_l(x,y)$ by
$
         \Psi_l(x,y)=F_l(x,y) + \sum_{l'}\int_{-1}^{+1}
         d\eta\,h_{l l'}(x,y,\eta)\,F_{l'}(x',y')
$
where
$
          x'=(\frac{1}{4}\,x^2+
    \frac{3}{4}\,y^2-\frac{\sqrt{3}}{2}\,xy\eta)^{1/2}\,,
         y'=(\frac{3}{4}\,x^2+
   \frac{1}{4}\,y^2+\frac{\sqrt{3}}{2}\,xy\eta)^{1/2}\,
$
and  $1 \leq{\eta}\leq 1$. The explicit form of the function
$h_{ll'}$ can be found in Refs.~\cite{MF,MGL}.
The functions $F_{l}(x,y)$ satisfy the boundary conditions
\begin{equation}
\label{BCStandardCor}
      F_{l}(x,y)\left.\right|_{x=0}
      =F_{l}(x,y)\left.\right|_{y=0}=0\,, \qquad
       \Psi_{l}(c,y)=0\,
\end{equation}
Note that the last of these conditions is a specific condition
corresponding to the hard-core model (see Ref.~\cite{KMS-JPB}).

Here we only deal with a finite number of
equations~(\ref{FadPartCor}), assuming that $l\leq l_{\rm max}$
where $l_{\rm max}$ is a certain fixed even number. The
condition $0\leq l\leq l_{\rm max}$ is equivalent to the
supposition that the potential $V(x)$ only acts in the two-body
states with $l=0,2,\ldots,l_{\rm max}$. The spectrum of the
Schr\"odinger operator for a system of three identical bosons
with such a potential is denoted by $\sigma_{3B}$.  We assume
that the potential $V(x)$ falls off exponentially and, thus,
$|V(x)|\leq C\exp(-\mu x)$ with some positive $C$ and $\mu$.
For the sake of simplicity we even assume sometimes that $V(x)$
is finite, i.\,e., \mbox{$V(x)=0$} for $x>r_0$, $r_0>0$.
Looking ahead, we note that, in fact, in our numerical
computations of the $^4$He$_3$ system at complex energies we
make a ``cutoff" of the interatomic He\,--\,He\,-\,potential at
a sufficiently large $r_0$.
The asymptotic
conditions as $\rho\rightarrow\infty$ and/or
$y\rightarrow\infty$ for the partial Faddeev components of the
$(2+1\rightarrow 2+1\,;\,1+1+1)$ scattering wave functions
for  $z=E+{\rm i}0$, $E>0$, read (see, e.\,g., Ref.~\cite{MF})  %
\begin{eqnarray}
\nonumber
 F_l(x,y;z) & = &
      \delta_{l0}\psi_d(x)\left\{\sin(\sqrt{z-\epsilon_d}\,y)
      + \exp({\rm i}\sqrt{z-\epsilon_d}\,y)
      \left[{\rm a}_0(z)+o\left(1\right)\right]\right\} \\
\label{AsBCPartS}
      && +
  \displaystyle\frac{\exp({\rm i}\sqrt{z}\rho)}{\sqrt{\rho}}
                \left[A_l(z,\theta)+o\left(1\right)\right]\,.
\end{eqnarray}
We assume that the $^4$He dimer has an only bound state with
an energy $\epsilon_d$, \mbox{$\epsilon_d<0$,} and wave function
$\psi_d(x)$, $\psi_d(x)=0$ for $0\leq x\leq c$. The notations $\rho$,
$\rho=\sqrt{x^2+y^2}$\,, and $\theta$, $\theta=\mathop{\rm
arctan}(y/x)$, are used for the hyperradius
and hyperangle. The coefficient ${\rm a}_0(z)$, $z=E+{\rm i}0$,
for $E>\epsilon_d$ is the elastic scattering amplitude.  The
functions $A_l(E+{\rm i}0,\theta)$ provide us, at $E>0$, the
corresponding partial Faddeev breakup amplitudes.
For real $z=E+{\rm i}0$, $E>\epsilon_d$, the
\mbox{$(2+1{\rightarrow}2+1)$} component of the $s$-wave partial
scattering matrix for a system of three helium atoms is given by
the expression
$$
{\rm S}_0(z)=1+2{\rm i}{\rm a}_0(z)\,.
$$
Our goal is to study the analytic continuation of the function
${\rm S}_0(z)$ into the physical sheet. As it follows from the
results of Refs.~\cite{MotMathNachr}, the function ${\rm
S}_0(z)$ is just that truncation of the total scattering matrix
whose roots in the physical sheet of the energy $z$ plane
correspond to the location of resonances situated in the
unphysical sheet adjoining the physical one along the spectral
interval $(\epsilon_d,0)$.

There are the following three important domains in
the the physical sheet.

1$^\circ$. The domain $\Pi^{(\Psi)}$ where the Faddeev components
$F_l(x,y;z)$ (and, hence, the wave functions $\Psi_l(x,y;z)$) can
be analytically continued in $z$ so that the differences
$
   \Phi_l(x,y;z)=F_l(x,y;z)-
   \delta_{l0}\psi_d(x)\sin(\sqrt{z-\epsilon_d}\,y)
$
at $z\in\Pi^{(\Psi)}\setminus\sigma_{3B}$
are square integrable.
This domain is described by the inequality
$$
\mathop{\rm Im}\sqrt{z-\epsilon_d}<{\rm min}\left\{\frac{\sqrt{3}}{2}\mu,\,
\sqrt{3}\sqrt{|\epsilon_d|}\right\}\,.
$$

2$^\circ$. The domain $\Pi^{(A)}$ where both the elastic
scattering amplitude ${\rm a}_0(z)$ and the Faddeev breakup
amplitudes $A_l(z,\theta)$ can be analytically continued in $z$,
$z\not\in\sigma_{3B}$, and where the continued functions
$F_l(x,y;z)$ still obey the asymptotic
formulas~(\ref{AsBCPartS}). This domain is described by the
inequalities
$$
\mathop{\rm Im}\sqrt{z}+\frac{1}{2}\mathop{\rm
Im}\sqrt{z-\epsilon_d} <
\frac{\sqrt{3}}{2}\sqrt{|\epsilon_d|}, \quad
\mathop{\rm Im}\sqrt{z}+\mathop{\rm Im}\sqrt{z-\epsilon_d} <
\frac{\sqrt{3}}{2}\mu\,.
$$

3$^\circ$. And finally, we distinguish the domain $\Pi^{(S)}$,
most interesting for us, where the analytic continuation in $z$,
$z\not\in\sigma_{3B}$, can be only done for the
amplitude ${\rm a}_0(z)$ (and consequently, for the
scattering matrix ${\rm S}_0(z)$); the analytic continuabilty of
the amplitudes $A_l(z,\theta)$ in the whole domain $\Pi^{(S)}$
is not required. The set $\Pi^{(S)}$ is a geometric locus
of points obeying the inequality
$$
\mathop{\rm Im}\sqrt{z-\epsilon_d}< \mathop{\rm
min}\left\{\frac{1}{\,\sqrt{3}\,}\,\sqrt{|\epsilon_d|},\,
  \frac{\,\sqrt{3}\,}{2} \mu\right\}\,.
$$

Since the spherical wave \mbox{$\exp({\rm
i}\sqrt{z}\,\rho)/\sqrt{\rho}$} in Eq.~(\ref{AsBCPartS}) is a
function rapidly decreasing in all the directions, the use of
the asymptotic condition (\ref{AsBCPartS}) is justified even if
$z\in\Pi^{(S)}\setminus\Pi^{(A)}$.  Outside of the domain
$\Pi^{(S)}$ the numerical construction of ${\rm S}_0(z)$ by
solving the Faddeev differential equations is, in general,
impossible.

\section{Numerical results}
\label{results}

In the present work  we search for the resonances of the $^4$He
trimer including the virtual levels as roots of ${\rm S}_0(z)$
and for the bound-state energies as positions of poles of ${\rm
S}_0(z)$.  All the results presented below are obtained for the
case $l_{\rm max}=0$.  In all our calculations,
\mbox{$\hbar^2/m=12.12$~K\,\AA$^2$.} As the interatomic
He\,--\,He\,-\,interaction we employed the HFD-B potential
constructed by R.\,A.\,Aziz and co-workers~\cite{Aziz87}.

The value of the core range $c$
is chosen to be so small that its further decrease
does not appreciably influence the dimer binding energy
$\epsilon_d$ and the trimer ground-state energy 
$E_t^{(0)}$. Unlike the paper~\cite{KMS-JPB}, where
$c$ was taken to be equal 0.7\,{\AA}, now we take
$c=1.3$\,{\AA}.  We have found that such a value of
$c$ provides at least six reliable figures of
$\epsilon_d$ and three figures of $E_t^{(0)}$.

Since the statements of Sect.~\ref{DiffFS} are valid, generally 
speaking, only for the potentials decreasing not slower than 
exponentially, we cut off the potential HFD-B setting $V(x)=0$ 
for $x>r_0$. We have established that this cutoff for 
\mbox{$r_0\gtrsim 95$\,{\AA}} provides the same values of 
$\epsilon_d$ ($\epsilon_d=-1.68541$\,mK), $E_t^{(0)}$ 
($E_t^{(0)}=-0.096$\,K) and scattering phases which were 
obtained in earlier calculations \cite{KMS-JPB} performed with 
the potential HFD-B. Also, we have found that the trimer 
excited-state energy \mbox{$E_t^{(1)}=-2.46$\,mK}.  Comparison 
of these results with results of other researchers can be found 
in Ref.~\cite{KMS-JPB}. In all the calculations of the present 
work we take \mbox{$r_0=100$\,{\AA}}.  Note that if the formulas 
from Sect. \ref{DiffFS} describing the holomorphy domains 
$\Pi^{(\Psi)}$, $\Pi^{(A)}$ and $\Pi^{(S)}$ are used for finite 
potentials, one should set in them $\mu=+\infty$.

A detailed description of the numerical method we use is 
presented in Ref.~\cite{KMS-JPB}.  When solving the 
boundary-value problem (\ref{FadPartCor}--\ref{AsBCPartS}) we 
carry out its finite-difference approximation in polar 
coordinates $\rho$ and $\theta$. In this work, we used the grids 
of dimension \mbox{$N_\theta=N_\rho=$\,600\,---\,1000}.  In 
essential, we chose the values of the cutoff hyperradius 
$\rho_{\rm max}=\rho_{N_\rho}$ from the scaling considerations 
(see~\cite{KMS-JPB}).  We solve the resultant 
block-three-diagonal algebraic system on the basis of the matrix 
sweep method. This allows us to dispense with writing the system 
matrix on the hard drive and to carry out all the operations 
related to its inversion immediately in RAM.  Besides, the 
matrix sweep method reduces almost by one order the computer 
time required for computations on the grids of the same 
dimensions as in~\cite{KMS-JPB}.

%
\begin{figure}
\centering
\epsfig{file=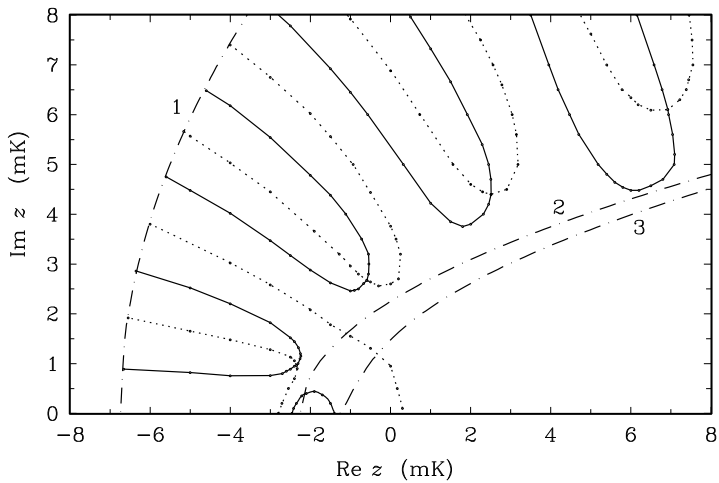,height=8.5cm}
\caption{Root locus curves of the real and imaginary parts of
the scattering matrix ${\rm S}_0(z)$. The solid lines correspond
to $\mathop{\rm Re} {\rm S}_0(z)=0$, while the tiny dashed
lines, to $\mathop{\rm Im} {\rm S}_0(z)=0$.  The numbers 1, 2
and 3 denote the boundaries of the domains $\Pi^{(\Psi)}$,
$\Pi^{(S)}$ and $\Pi^{(A)}$, respectively. Complex roots of the
function ${\rm S}_0(z)$ are represented by the intersection
points of the curves $\mathop{\rm Re} {\rm S}_0(z)=0$ and
$\mathop{\rm Im} {\rm S}_0(z)=0$ and are located at
\mbox{$(-2.34+{\rm i}\,0.96)$}\,mK, \mbox{$(-0.59+{\rm
i}\,2.67)$}\,mK, \mbox{$(2.51+{\rm i}\,4.34)$}\,mK and
\mbox{$(6.92+{\rm i}\,6.10)$}\,mK.}
\label{RootLines}
\end{figure}

Because of the symmetry relationship $\overline{{\rm
S}_0(z)}={{\rm S}_0(\overline{z})}$ we performed all the
calculations for ${\rm S}_0(z)$ only at $\mathop{\rm Im} z\geq
0$.  First, we calculated the root lines of the functions
$\mathop{\rm Re}{\rm S}_0(z)$ and $\mathop{\rm Im}{\rm S}_0(z)$.
For the case of the grid parameters \mbox{$N_\theta=N_\rho=600$}
and \mbox{$\rho_{\rm max}=600$}\,{\AA} these lines are depicted
in Fig.~1. Both resonances (roots of ${\rm S}_0(z)$) and
bound-state energies (poles of ${\rm S}_0(z)$) of the $^4$He
trimer are associated with the intersection points of the curves
$\mathop{\rm Re}{\rm S}_0(z)=0$ and $\mathop{\rm Im}{\rm
S}_0(z)=0$.  In Fig.~1, along with the root lines we also plot
the boundaries of the domains $\Pi^{(S)}$, $\Pi^{(A)}$ and
$\Pi^{(\Psi)}$.  One can observe that the ``good" domain
$\Pi^{(S)}$ includes none of the points of intersection of the
root lines $\mathop{\rm Re}{\rm S}_0(z)=0$ and $\mathop{\rm
Im}{\rm S}_0(z)=0$.  The caption for Fig.~1 points out positions
of four ``resonances", the roots of ${\rm S}_0(z)$, found
immediately beyond the boundary of the domain $\Pi^{(S)}$.  It
is remarkable that the ``true" (i.\,e., getting inside
$\Pi^{(S)}$) virtual levels and then the energies of the excited
(Efimov) states appear just due to these (quasi)resonances when
the potential $V(x)$ is weakened.

Following~\cite{Gloeckle}--\cite{KMS-JPB},
instead of the initial potential $V(x)=V_{\rm HFD-B}(x)$,
we, further,  consider the potentials
$
          V(x)=\lambda\cdot V_{\rm HFD-B}(x).
$
To establish the mechanism of formation of new excited states in
the $^4$He trimer, we have first calculate the scattering
matrix ${\rm S}_0(z)$ for $\lambda<1$.
\begin{table}
\label{tableTrimerVirtEsc}
\caption
{Dependence of the dimer binding energy $\epsilon_d$ and the
differences $\epsilon_d-E_t^{(1)}$, $\epsilon_d-E_t^{(2)}$, 
$\epsilon_d-E_t^{(2)*}$ and $\epsilon_d-E_t^{(2)**}$ (all in mK)
between this energy and the trimer exited-state energies
$E_t^{(1)}$, $E_t^{(2)}$ and the  virtual-state energies
$E_t^{(2)*}$, $E_t^{(2)**}$ on the factor $\lambda$.}
\centering
\begin{tabular}{|c|c|c|c|c|c|}
\hline\hline
 $\lambda$   & $\epsilon_d$    & $\epsilon_d-E_t^{(1)}$ &
  $\epsilon_d-E_t^{(2)*}$ & $\epsilon_d-E_t^{(2)**}$ &
  $\epsilon_d-E_t^{(2)}$ \\
\hline
 0.995 & $-1.160$ &  0.710       & --           & --       & -- \\
 0.990 & $-0.732$ &  0.622       & --           & --       & -- \\
0.9875 & $-0.555$ &  0.573       & 0.473        & 0.222    & -- \\
 0.985 & $-0.402$ &  0.518       & 0.4925       & 0.097    & -- \\
 0.980 & $-0.170$ & 0.39616      & 0.39562      & 0.009435 & -- \\
 0.975 & $-0.036$ & 0.2593674545 & 0.2593674502 & --  & 0.00156 \\
\hline\hline
\end{tabular}
\end{table}
In Table~4.1 for some values of $\lambda$ from the interval
between 0.995 and 0.975, we present the positions of roots and
poles of ${\rm S}_0(z)$, we have obtained at real
$z<\epsilon_d(\lambda)$. We have found that for a value of
$\lambda$ slightly smaller than $0.9885$, the (quasi)resonance
closest to the real axis (see Fig.~1) gets on it and transforms
into a virtual level (the root of ${\rm S}_0(z)$) of the second
order corresponding to the energy value where the graph of ${\rm
S}_0(z)$, $z\in{\Bbb R}$, $z<\epsilon_d$, is tangent to the axis
$z$. This virtual level is preceded by the (quasi)resonances
\mbox{$z=(-1.04+{\rm i}\,0.11)$}\,mK for $\lambda=0.989$ and
\mbox{$z=(-0.99+{\rm i}\,0.04)$}\,mK for $\lambda=0.9885$.  With
a subsequent decrease of $\lambda$ the virtual level of the
second order splits into a pair of the first order virtual
levels $E_t^{(2)*}$ and $E_t^{(2)**}$, $E_t^{(2)*}<E_t^{(2)**}$
which move in opposite directions. A characteristic behavior of
the scattering matrix ${\rm S}_0(z)$ when the resonances
transform into virtual levels is shown in Fig.~2. The virtual
level $E_t^{(2)**}$ moves towards the threshold $\epsilon_d$ and
``collides'' with it at $\lambda<0.98$. For $\lambda=0.975$ the
function ${\rm S}_0(z)$ has no longer the root corresponding to
$E_t^{(2)**}$. Instead of the root, it acquires a new pole
corresponding to the second excited state of the trimer with the
energy $E_t^{(2)}$.  We expect that the subsequent Efimov levels
originate from the virtual levels just according to the same
scheme as the level $E_t^{(2)}$ does.

%
\begin{figure}
\centering
\epsfig{file=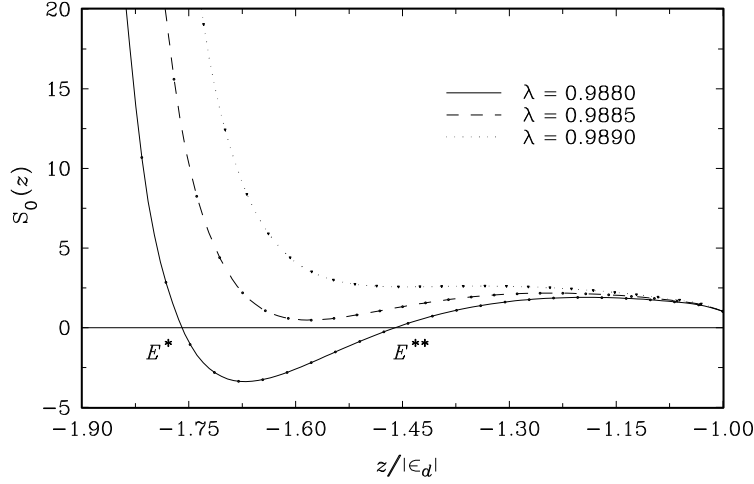,height=8.5cm}
\caption{Graphs of the function ${\rm S}_0(z)$ at real
$z\leq\epsilon_d$  for three values of $\lambda<1$ .
The notations used: $E^{*}=E_{t}^{(2)*}/|\epsilon_d|$,
$E^{**}=E_{t}^{(2)**}/|\epsilon_d|$.}
\label{S0Virt}
\end{figure}

The other purpose of the present investigation is to determine
the mechanism of disappearance of the excited state of the
helium trimer when the two-body interactions become stronger
owing to the increasing $\lambda>1$.  It turned out that this
disappearance proceeds just according to the scheme of the
formation of new excited states; only the order of occurring
events is inverse.  The results of our computations of the
energy $E_t^{(1)}$ when $\lambda$ changes from 1.05 to 1.17 are
given in Table~4.2.
\begin{table}
\label{tableTrimerExcitedVirtual}
\caption
{Dependence of the dimer energy $\epsilon_d$, the difference
$\epsilon_d-E_t^{(1)}$ between $\epsilon_d$ and the trimer
exited-state energy  $E_t^{(1)}$ and the difference
$\epsilon_d-E_t^{(1)*}$ between $\epsilon_d$ and the trimer
virtual-state energy $E_t^{(1)*}$ (all in mK) on the factor
$\lambda$.}
\centering
\begin{tabular}{|c|c|c||c|c|c|}
\hline\hline
 $\lambda$ & $\epsilon_d$ & $\epsilon_d-E_t^{(1)}$ &
 $\lambda$ & $\epsilon_d$ & $\epsilon_d-E_t^{(1)*}$ \\
\hline
1.05 & $-12.244$ & 0.873 & 1.18 & $-82.927$  &  0.001  \\
1.10 & $-32.222$ & 0.450 & 1.20 & $-99.068$  &  0.057  \\
1.15 & $-61.280$ & 0.078 & 1.25 & $-145.240$ &  0.588  \\
1.16 & $-68.150$ & 0.028 & 1.35 & $-261.393$ &  3.602  \\
1.17 & $-75.367$ & 0.006 & 1.50 & $-490.479$ &  12.276 \\
\hline\hline
\end{tabular}
\end{table}
In the interval between $\lambda=1.17$ and $\lambda=1.18$ there
occurs a ``jump" of the level $E_t^{(1)}$ on the unphysical
sheet and it transforms from the pole of the function ${\rm
S}_0(z)$ into its root, $E_t^{(1)*}$, corresponding to the
trimer virtual level.  The results of calculation of this
virtual level where $\lambda$ changes from 1.18 to 1.5 are
also presented in Table~4.2.

More details of our techniques and material presented
will be given in an extended article~\cite{KolMotYaF}.
\begin{acknowledge}
The authors are grateful to Prof.~V.\,B.\,Belyaev and 
Prof.~H.\,Toki for help and assistance in calculations at the 
supercomputer of the Research Center for Nuclear Physics of the 
Osaka University, Japan.  One of the authors (A.\,K.\,M.) is 
much indebted to Prof.~W.\,Sandhas for his hospitality at the 
Universit\"at Bonn, Germany. The support of this work by the 
Deutsche Forschungsgemeinschaft and Russian Foundation for Basic 
Research is gratefully acknowledged.
\end{acknowledge}


\SaveFinalPage

\begin{thebibliography}{99}
\bibitem{MotMathNachr} A. K. Motovilov:
       Math. Nachr. {\bf 187}, 147 (1997) 
       (LANL E-print {\tt funct-an/9509003})
\bibitem{Gloeckle}
       T. Cornelius, W. Gl\"ockle:
       J. Chem. Phys. {\bf 85}, 3906 (1986)
\bibitem{EsryLinGreene}
     B. D. Esry, C. D. Lin, C. H. Greene:
    Phys. Rev.~A. {\bf 54}, 394 (1996)
\bibitem{KMS-JPB}  E. A. Kolganova,  A. K. Motovilov,  S. A. Sofianos:
    J.~Phys. B. {\bf 31}, 1279 (1998) 
    (LANL E-print {\tt physics/9612012})
\bibitem{Nielsen}
     E. Nielsen,  D. V. Fedorov,  A. S. Jensen:
     LANL E-print {\tt physics/9806020}
\bibitem{KartavtsevPenkov} O. I. Kartavtsev, F. M. Penkov: 
     In: {\it 16th European Conference on Few-Body Problems 
     in Physics (Autrans, 1\,--\,6~June 1998), Abstract Booklet}, 
     p.~137. Grenoble 1998
\bibitem{TFDA} L. Tomio, T. Frederico, A. Delfino, A. E. A. Amorim: 
     {\it Ibid.}, p.~150.  
\bibitem{VEfimov}
      V. Efimov: Nucl. Phys. A. {\bf 210}, 157 (1973)
\bibitem{Aziz87}
      R. A. Aziz,  F. R. W. McCourt,  C. C. K. Wong:
      Mol. Phys. {\bf 61}, 1487 (1987)
\bibitem{MF}
    L. D. Faddeev, S. P. Merkuriev:
    {\it Quantum Scattering Theory
    for Several Particle Systems}.
    Doderecht: Kluwer Academic Publishers 1993
\bibitem{MGL}
      S. P. Merkuriev, C. Gignoux, A. Laverne:
     Ann. Phys. (N.Y.)  {\bf 99}, 30 (1976)
\bibitem{KolMotYaF} E. A. Kolganova, A. K. Motovilov:
     Preprint JINR E4-98-243 (LANL E-print {\tt physics/9808027}).
\end{thebibliography}
\end{document}